\documentclass[a4paper,11pt,dvips]{article}
cm \voffset = -2truecm

\usepackage{graphicx}
\usepackage{amsmath}
\usepackage{amssymb}
\usepackage{latexsym}

\begin{document}
\begin{titlepage}
\setcounter{page}{1}
\renewcommand{\thefootnote}{\fnsymbol{footnote}}

\begin{flushright}
\end{flushright}

\vspace{5mm}
\begin{center}

 {\Large \bf Bipartite and Tripartite Entanglement of Truncated \\ Harmonic
Oscillator Coherent States
 via Beam Splitters}

\vspace{5mm}

{\bf M. Daoud$^{a,b}$\footnote {daoud@pks.mpg.de,
m$_-$daoud@hotmail.com}}, {\bf A.
Jellal$^{c,d,e}$\footnote{ajellal@ictp.it, jellal@ucd.ma}}, {\bf
E.B. Choubabi$^{d,e}$\footnote{choubabi@gmail.com}}
 and
{\bf E.H. El Kinani$^{f}$\footnote{elkinani@gmail.com}}

\vspace{5mm}

{$^{a}$\em  Max Planck Institute for Physics of Complex Systems,
 D-01187 Dresden, Germany}

{$^{b}$\em Department of Physics,
Faculty of Sciences, Ibn Zohr University},\\
{\em PO Box 8106, 80006 Agadir, Morocco}

{$^c$\em Physics Department, College of Sciences, King Faisal University,\\
PO Box 380, Alahsa 31982, Saudi Arabia}

{$^d$\em Saudi Center for Theoretical Physics, Dhahran, Saudi
Arabia}

{$^{e}$\em Theoretical Physics Group,  
Faculty of Sciences, Choua\"ib Doukkali University},\\
{\em PO Box 20, 24000 El Jadida, Morocco}

{$^f$\em Mathematics Department, Faculty of Science and Technical, Moulay Ismail University,\\
 P.O. Box 509, Errachidia, Morocco}

\vspace{3cm}

\begin{abstract}

We introduce a special class of truncated Weyl--Heisenberg algebra
and discuss the corresponding Hilbertian and analytical
representations. Subsequently, we study the effect of a quantum
network of beam splitting on coherent states of this nonlinear class
of harmonic oscillators. We particularly focus on quantum networks
involving one and two beam splitters and examine the degree of
bipartite as well as tripartite entanglement using the linear
entropy.

\end{abstract}
\end{center}
\end{titlepage}

\newpage

\section{Introduction}

Over the past few years, the entanglement of quantum systems was
realized to be a valuable and crucial  resource in quantum
information processing. It allows for powerful new communication and
computational tasks that are not possible classically. One may quote
for instance quantum teleportation \cite{Ben1}, superdense coding
\cite{Ben2}, quantum key distribution \cite{Eckert}, telecloning
\cite{Murao}, quantum cryptography \cite{Fuchs} and quantum
computation \cite{Rausschendorf,Gottesman}. Then it is not
surprising then that over the last decade much efforts has been
devoted  to capture, quantify and assess the power of quantum
entanglement. In this sense, many authors studied the development of
a quantitative theory of entanglement and the definition of its
basic measure (concurrence, entanglement of formation and linear
entropy \cite{Rungta,Ben3,Wootters,Coffman}). Entangled quantum
systems can exhibit correlations that cannot be explained on the
basis of classical laws and the entanglement in a collection of
states is clearly a signature of non-classicality \cite{Markham}.

In quantum theory, the states minimizing the quantum fluctuations,
which are closest to the classical ones are the coherent states
\cite{Klauder,Perelomov}. This motivated  the considerable interest
in the entanglement of coherent states
\cite{Sanders,Gerry1,Luis,Wang}. The coherent state approach is not
just a mathematical tool, but it my also helps to understand the
entanglement for such states, which provide a bridge between quantum
and classical worlds. It is also important to stress that in
connection with quantum entanglement, many experimental results were
obtained. In fact, the experimental generation and characterization
of the entangled electromagnetic states can be achieved using type I
or type II parametric down conversion \cite{Kwiat}. Another
experimentally accessible device, which can be used to generate
optical  entangled states, is the beam splitter
\cite{Tan,Sanders2,Paris}. In quantum optics  the action of a beam
splitter, which is essentially a coupling of two electromagnetic
modes, can be represented by a unitary operator relating the input
and the output states. In general, the output state is a
superposition of the Fock states which is entangled, except the
harmonic oscillator coherent states. Indeed, the harmonic oscillator
coherent states does not exhibit entanglement when passed through
one arm of 50:50 beam splitter while the second arm is left in the
empty vacuum state \cite{Kim}. In the same sense, the entanglement
behavior of the $SU(2)$ spin coherent states, when passed through a
beam splitter, was examined in \cite{Markham}. Similar study was
done in \cite{Gerry2} for the $SU(1,1)$ coherent states.

On the other hand, the truncated harmonic oscillator was used by
Pegg and Barnett \cite{Pegg} to define the phase states for the
quantized single modes of the electromagnetic field. They suggested
to truncate to some finite (but arbitrarily large) order the
infinite-dimensional representation space of the oscillator algebra.
This was done to get rid of the difficulty related to the
infinite-dimensional character of the representation space of the
Weyl--Heisenberg algebra,  which constitutes a drawback in defining
a phase operator in a consistent way
\cite{Louisell,Susskind,Carruthers}. Motivated by these
investigations and in particular~\cite{Pegg}, we propose a refined
version of the truncated oscillator algebra introduced by Pegg and
Barnett. More precisely, we introduce a  nonlinear class of
Weyl--Heisenberg algebra and analyze its representations. We
construct the associated coherent states using a suitable analytical
realization and we study the degree of entanglement of such states
when passed through a quantum network of beams splitters.

The outline of the paper is as follows. In section 2, we introduce a
generalized Weyl--Heisenberg algebra. We discuss the corresponding
Hilbertian representation and the analytical Bargmann realization.
It is remarkable that the obtained analytical realization provides
us with an over-complete set of states and then constitute a system
of coherent states in the
 Klauder--Perelomov sense. In section 3, we define the action
of a quantum network of $k$ beam splitters. We show that this action
leads to the $SU(k+1)$ coherent states labeled by complex variables
and related to the reflection and transmission parameters of beam
splitters. Based on this result, we investigate in detail the effect
of one and two beam splitters on the generalized Weyl--Heisenberg
algebra coherent states. We derive the linear entropy to study the
bipartite entanglement degree of the output states. study the
entanglement in a system of three particles. Concluding remarks
close this paper.

\section{Weyl--Heisenberg algebra and Fock--Bargmann realization} 

We start by introducing and discussing some interesting properties
of a generalized version of the truncated harmonic oscillator, which
involves a half positive integer $s$.  In the limiting case when
this parameter goes to the infinity, one recovers the usual harmonic
oscillator with infinite dimensional Fock space. This provides us
with another mathematical tool to deal with truncation of the
ordinary harmonic oscillator that is different from one discussed in
\cite{Pegg}. We mention that the idea of truncated Weyl--Heisenberg
is mainly inspired by the polynomial deformation of Lie algebras
introduced in \cite{Higgs,Sklyanin} and extensively discussed in the
context of quantum algebras \cite{Rocek,Bonatsos,Abdesselam}. The
truncated Weyl--Heisenberg algebra discussed here can be viewed as a
special case of $f$-deformed oscillators introduced in
\cite{Manko1,Manko2} which is relevant in the algebraic  description
of a large class of nonlinear quantum systems.

\subsection{Finite dimensional Weyl--Heisenberg algebra}

The basic ingredient that we shall use in what follows is the
generalized Weyl--Heisenberg algebra characterized by a positive
real parameter $(s > 0)$. This algebra is generated by the set of
operators $\{a^+ , a^- ,N , ~{\mathbb I}\}$ satisfying the relations
\begin{equation}
[N, a^-] = -a^-,\qquad [N, a^+ ] = + a^+,\qquad
 [a^- , a^+] =  {\mathbb I} - \frac{N}{s}
\label{algebra}
\end{equation}
where $N$ is the number operator 
and the element ${\mathbb I}$ commutes with all other operators.
Clearly, when $s \to \infty$, end up with the ordinary
Weyl--Heisenberg algebra. In this respect, the parameter $s$ can be
regarded as a measure of the distortion of this algebra. For
convenience, we assume that $2s \in {\mathbb N}$. The generalized
oscillator algebra can be naturally represented, on the Fock space
${\cal F}$  of the eigenstates of the number operator $N$, by
\begin{equation}
N~\vert n \rangle ~= ~n ~\vert n \rangle , \qquad \langle
n|m\rangle=\delta_ {nm}, \qquad  n,m \in \mathbb{N}
\end{equation}
and the vacuum state satisfies $a^- \vert 0 \rangle =0$. We define
the actions of the operators $a^-$ and $a^+$ on ${\cal F}$ as
\begin{equation}
 a^-\vert n \rangle =\sqrt{F(n)}\vert n-1 \rangle ,\qquad
 a^+\vert n \rangle=\sqrt{F((n+1)}\vert n+1\rangle.
\label{action}
\end{equation}
Note that,  $a^+$ and $a^-$ are mutually adjoint, namely $a^+ =
(a^-)^{\dagger}$ and  $N$ is, in general, different from the product
$a^+a^-$. The structure function $F(\cdot)$ is an analytic function
with the properties $F(0)=0$ and $F(n)>0$, $n=1,\cdots$. $F(\cdot)$
is characteristic to the distortion or the truncation scheme and
satisfies the following recursion formula
\begin{equation}
F(n+1) - F(n) = 1 - \frac{n}{s}
\end{equation}
which gives by simple iteration the form
\begin{equation}
F(n) = \frac{n}{2s} (2s+1 - n).
\end{equation}
Note that, the structure function obeys the condition
\begin{equation}
 F(2s+1)=0
\end{equation}
and the creation-annihilation operators satisfy the nilpotency
relations
\begin{equation}
(a^-)^{2s+1}= 0, \qquad (a^+)^{2s+1}=0.
\end{equation}
 This
means that the corresponding representation is $(2s+1)$-dimensional.
As evoked above the truncated oscillator algebra (\ref{algebra})
constitutes a particular variant of the $f$-deformed oscillators
\cite{Manko1,Manko2}. Indeed, one has
\begin{equation}
a^- = b^-~f(N), \qquad a^+ = f(N)~ b^+, \qquad N =
b^+b^-,\label{realization}
\end{equation}
where the function $f(N)$, reflecting the distortion from the usual
bosons, is given by
$$ f(N) = \sqrt{1-\frac{N - 1}{2s}}.$$
The equation (\ref{realization}) is a Holstein-Primakoff realization
of the algebra (\ref{algebra}). Finally, we point out that for $s$
large, we have
\begin{equation}
a^{\pm} \sim b^{\pm}
\end{equation}
traducing that the distorted algebra (\ref{algebra}) coincides with
the linear harmonic oscillator one. More importantly, the
generalized Weyl--Heisenberg algebra provides us with a simply
framework to deal with finite dimensional oscillator system. It is
important to stress that this algebra is similar to one introduced
in \cite{Daoud} in order to define the phase operator for nonlinear
oscillators and to derive the associated temporally stable phase
states. In this respect, the generalized Weyl--Heisenberg algebra
can be viewed as a refined version of truncated harmonic oscillator.

Of course the problem of practical realization of the  finite
dimensional nonlinear quantum oscillator algebra arises. At this
point one should emphasize that many exactly solvable systems enjoy
the generalized Weyl--Heisenberg symmetry and are possessing finite
dimensional representation space. One may quote for instance one
dimensional quantum systems having finite discrete spectrums like
ones evolving in the modified P\"oschl-Teller \cite{Recamier1} and
Morse \cite{Daoud2} potentials, see also \cite{Daoud3}. Quantum
systems described by a nonlinear Hamiltonian are familiar  in the
context of nonlinear quantum optics as for instance electromagnetic
field propagating trough a nonlinear Kerr medium \cite{Klimov1}.
Indeed, the dynamical evolution of the electromagnetic field
propagating in a Kerr medium can be described by the following
Hamiltonian
$$ H_{\rm Kerr} = b^+b^- ~- ~\kappa ~ b^{+2}b^{-2}$$
where $\kappa$ characterizes the Kerr nonlinearity. Setting $\kappa
= \frac{1}{2s}$, the Hamiltonian $H_{\rm Kerr}$ coincides with the
function structure $F(N)$ with $N \equiv b^+b^-$. It is also
interesting to stress that the ladder operators $ a^+$ and $a^-$ can
be related to Stokes operators introduced in \cite{Luis1,Luis2} in
order to define a unitary operator representing the exponential of
the phase difference between two modes of the electromagnetic field.
This relation can be expressed as
$$ s_+ = \sqrt{s}~ a_+, \qquad s_- = \sqrt{s}~ a_-, \qquad s_3 = \frac{1}{2}\left(N - s\right)$$
and one can simply verify that the Stokes generators $ s_+ $, $ s_-$
and $s_3$ satisfy the commutation relations of the $su(2)$ algebra.

We close this subsection by noting that, in recent years, a special
interest is devoted to the possibility of generating and
manipulating systems, whose dynamics can be closed within a finite
set of $n$-photon states. Such systems are very important for the
implementation of models in the quantum information theory. The
experimental devices leading to the finite-dimensional state
generation are referred, in the literature,  as linear
\cite{Pegg1,Barnett} or non-linear \cite{Leo1,Leo2} quantum
scissors.

\subsection{Fock-Bargmann realization}

The second ingredient needed for our task is the coherent states
associated with the above generalized algebra. A simply way to
construct these states is to use the analytical Bargmann
representation. Indeed, we realize the annihilation operator
\begin{equation}
a^-\longrightarrow \frac{d}{dz}
\end{equation}
as derivation with respect to a complex variable $z$. The elements
of the Fock space are realized as follows
\begin{equation}
| n \rangle  \longrightarrow C_{n} z^{n}. \label{coresp}
\end{equation}
Using the action of the operator $a^-$ on the Fock space ${\cal F}$
given by (\ref{action}) and the correspondence (\ref{coresp}), one
obtains the following recursion formula
\begin{equation}
(n+1) C_{n+1}= \sqrt{F(n+1)} C_{n}.
\end{equation}
It follows that the coefficients $C_n$ are given by
\begin{equation}
C_n = C_0 \sqrt{\frac{2s!}{(2s)^n n! (2s-n)!}}.
\end{equation}
To simplify, we set hereafter $C_0 = 1$. Having the expression of
the coefficients $C_n$, one can determine the differential action of
the
 creation operator $a^+$. Indeed, since the operator $N$ acts as
\begin{equation}
N \longrightarrow z\frac{d}{dz}
\end{equation}
one can easily see, using the action of the generator $a^+$ on the
Fock space,  that
\begin{equation}
a^+ \longrightarrow z - \frac{z^2}{2s}\frac{d}{dz}.
\end{equation}
On the other hand, a general vector of ${\cal F}$
$$|\phi \rangle =
\sum_{n = 0}^{2s} \phi_{n}| n \rangle$$ is represented in the
Bargmann space as
\begin{equation}
\phi(z) = \sum_{n=0}^{2s}\phi_{n} C_{n} z^{n}.
\end{equation}
The inner product of two functions $\phi$ and $\phi'$ is defined now
as
\begin{equation}
\langle\phi'|\phi\rangle = \int  d^2z \Sigma(z){\overline{\phi'(z)}}
\phi(z)
\end{equation}
where the integration is carried out on the whole complex plane. The
integration measure $\Sigma$, assumed to be isotropic, can be
computed by choosing $|\phi\rangle = | n \rangle $ and
$|\phi'\rangle = | n' \rangle $. Thus, one has to look for a
solution of the following moment equation
\begin{equation}
 2\pi \int_{0}^{+\infty} d\varrho \Sigma(\varrho)
\varrho^{2n+1} = \frac{(2s)^n n!(2s-n)!}{(2s)!} \label{integral}
\end{equation}
where $\varrho  = |z|$. To find the function satisfying the equation
(\ref{integral}), we use Mellin transform technique to obtain
\begin{equation}
\Sigma (\varrho) = \frac{2s+1}{\pi} ( 1 +
\frac{\rho^2}{2s})^{-2s-2}.\label{measure}
\end{equation}
At this stage, one can write the function $\phi(z)$ as the product
of the state $|\phi\rangle $ with some ket $| \bar z \rangle$
labeled by the complex conjugate of the variable $z$. This is
\begin{equation}
\phi(z)=  {\cal N}\langle  \bar z |\phi \rangle.
\end{equation}
Taking $|\phi\rangle = | n \rangle$, we obtain
\begin{equation}
| z \rangle =  {\cal N} \sum_{n = 0}^{2s} \sqrt{\frac{(2s)!}{(2s)^n
n!(2s-n)!}} z^n | n \rangle. \label{cs}
\end{equation}
 The normalization constant for the
the states (\ref{cs}) is given by
\begin{equation}
{\cal N}= \bigg( 1 + \frac{|z|^2}{2s}\bigg)^{-s}. \label{norme}
\end{equation}

It is important to remark that the states (\ref{cs}) are coherent in
the Klauder--Perelomov sense \cite{Klauder,Perelomov}. Indeed, it is
easy to see that the states $\vert z \rangle$ can be written as
$$\vert z \rangle = {\cal N} \exp(za^+) \vert 0 \rangle.$$
They satisfy the over-completeness property
$$\int d^2z \Sigma(\vert z \vert) \vert z \rangle \langle z \vert = \sum_{n=0}^{2s} \vert n \rangle \langle n \vert$$
where the measure $\Sigma(\vert z \vert)$ is given by
(\ref{measure}). It should be noticed the states (\ref{cs}) are
similar to nonlinear coherent states derived  in \cite{Recamier2}.
They coincide with the standard Glauber coherent states for the
usual harmonic oscillator at large $s$.

\section{ Quantum network of beam splitters}

In the recent years, the study of entangled states  has revived
interest in the beam splitter. This is due to the fact that this
device offers a simple way to probe the quantum nature of
electromagnetic field through simple experiments.  As mentioned
before, many authors studied the behavior of quantum states when
passed through a beam splitter. This device is an optical element
with two input ports and two output ports that, in some sense,
governs the interaction  of two harmonic oscillators. The input and
output boson operators are related by a unitary transformation which
is an element of the $SU(2)$ group. Recently, a quantum network of
beam splitters was used to create multiparticle entangled states of
continuous variables \cite{Van} and also multiparticle entangled
coherent states \cite{Wang2}. Here we examine the entanglement of
the coherent states (\ref{cs}) passing through a quantum network of
beam splitters. Note that,  a set of $k$ beam splitters can be
experimentally used to generate $SU(k+1)$ coherent states labeled by
the reflection and transmission parameters of the involved beam
splitters.

\subsection{ Generation of $SU(k+1)$ coherent states}

 The
most general transformation defining a quantum network of $k$ beam
splitters is given by the unitary operator
\begin{equation}
{\cal B}_k = {\cal B}_{k,k+1}(\theta_k)  {\cal
B}_{k-1,k}(\theta_{k-1}) \cdots {\cal B}_{1,2}(\theta_{1})
\label{operatorB}
\end{equation}
 where  the operators
\begin{equation}
{\cal B}_{l,l+1}(\theta_l) = \exp\left[\frac{i}{2} \theta_l
\left(b_l^+b_{l+1}^- + b_l^-b_{l+1}^+\right)\right]
\end{equation}
are the ordinary $SU(2)$  beam splitters with $l = 1 , 2, \cdots,
k$. The objects $b^+_l$ and $b^-_l$ are the usual harmonic
oscillator ladder operators. The reflection and transmission
coefficients
\begin{equation}
 t_l = \cos\frac{\theta_l}{2}, \qquad  r_l = \sin\frac{\theta_l}{2}.
\end{equation}
are defined  in terms of the angles $\theta_l$. The operator ${\cal
B}_k$ is actually acting on the states $\vert n_1 , n_2, \cdots ,
n_k \rangle$  of the usual $k$-dimensional harmonic oscillator. If
the input state is  $\vert n_1 , n_2, \cdots , n_k \rangle$, then
the  ${\cal B}_k $ action leads to the following Fock states
superposition
\begin{equation}
{\cal B}_k \vert n_1 , n_2, \cdots , n_{k+1} \rangle =
\sum_{m_1,m_2,\cdots,m_{k+1}} {\cal B}_{n_1 , n_2, \cdots , n_{k+1}
}^{m_1,m_2,\cdots,m_{k+1}}
 \vert m_1 , m_2, \cdots , m_{k+1} \rangle
\end{equation}
and in general the output is a $(k+1)$-particle entangled state. On
the other hand, the action of the unitary operator ${\cal B}_k$ on
the state $\vert n_1 , 0, \cdots , 0 \rangle$ gives
\begin{eqnarray}
{\cal B}_k \vert n_1 , 0, \cdots , 0 \rangle &=& {\cal C}
\sum_{n_2=0}^{n_1}\sum_{n_3=0}^{n_2} \cdots \sum_{n_{k+1}=0}^{n_{k}}
\frac{ \xi_{1}^{n_2}\xi_{2}^{n_3}\cdots \xi_{k}^{n_{k+1}}\sqrt{n_1!}}{\sqrt{(n_1 - n_2)!(n_2-n_3)!\cdots(n_{k}-n_{k+1})!n_{k+1}!}} \nonumber\\
&& \times \vert n_1-n_2 , n_2-n_3, \cdots , n_{k+1} \rangle
\label{suncs}
\end{eqnarray}
where the normalization constant reads as
\begin{equation}
 {\cal C}  =  \left(1 +  |\xi_{1}|^2 + |\xi_{1}|^2|\xi_{2}|^2\cdots +|\xi_{1}|^2|\xi_{2}|^2\cdots |\xi_{k}|^2\right)^{-{n_1}/{2}}
\end{equation}
and the new variables $\xi$
\begin{equation}
\xi_l = i t_{l+1}~\frac{r_l}{t_l}\ \qquad {\rm for} \ (l =
1,2,\cdots,k-1), \qquad \xi_{k} = i \frac{r_{k}}{t_{k}}
\end{equation}
are defined in terms of the reflection and transmission coefficients
of the beam splitters constituting the network. Then, the output
state (\ref{suncs}) turns out to be the $SU(k+1)$ coherent state
associated with the completely symmetric representation labeled by
the integer $n_1$, see for instance \cite{Daoud1} where similar
notations were used.  This means that a network of $k$ beam
splitters can be used as experimental device to generate $SU(k+1)$
coherent states.

\subsection{ Bipartite entanglement of truncated harmonic oscillator coherent states}

We start by investigating the effect of one beam splitter on the
coherent states (\ref{cs}) associated with the generalized or
truncated harmonic oscillator introduced in the first section. As
discussed above, for $s$ large, the dimension of the generalized
oscillator Fock space becomes infinite. In this case one can easily
check that the coherent states (\ref{cs}) reduce to Glauber coherent
states of the infinite dimensional harmonic oscillator. It has been
proven that the Glauber states are the only pure states that when
passed through one arm of the beam splitter, the output resultant
states are disentangled \cite{Aharonov}. In this respect, the main
task here is to determine the behavior of the entanglement in terms
of the Fock space dimension and to prove that the entanglement
disappears when the Fock space dimension goes to infinity. For that
end, we consider the situation where the coherent state (\ref{cs})
is injected into one port while no photon is injected into the other
port.

Using the equation (\ref{suncs}) for $k=1$, it is simply verified
that the action of the beam splitter on the input state $\vert z , 0
\rangle$ can be cast in the following form
\begin{equation}
{\cal B}_{1,2}(\theta_1) \vert z , 0 \rangle = {\cal N}
\sum_{n=0}^{2s} \sum_{q=0}^{2s-n}
\sqrt{\frac{(2s)!}{(2s-n-q)!n!q!}}~ t_1^n (ir_1)^q
~\frac{z^{q+n}}{(\sqrt{2s})^{q+n}} ~\vert n , q \rangle
\label{actionB12}
\end{equation}
where ${\cal N}$ is given by (\ref{norme}). To determine the degree
of entanglement of the beam splitter output sates, we use the linear
entropy which is the upper bound of the Von Neumann entropy
\cite{Markham}. This is 
\begin{equation}
S = 1 - {\rm Tr}(\rho_1^2)
\end{equation}
where $\rho_1 = {\rm Tr}_2 \rho_{12}$ is the reduced density
operator obtained by tracing over the states of the second system.
Using (\ref{actionB12}), the linear entropy is obtained as
\begin{equation}
S = 1 - \sum_{n=0}^{2s} \sum_{n'=0}^{2s} \sum_{q=0}^{{\rm min}(2s-n,
2s-n')}  \sum_{q'=0}^{{\rm min}(2s-n, 2s-n')} b(n,n',q)b(n',n,q')
\end{equation}
where
\begin{equation}
b(n,n',q) = \vert {\cal N} \vert^2 (2s)! \frac{|z|^{2q}}{(2s)^q q!}
~ t^{n+n'} r^{2q} ~\frac{z^{n}\bar z^{n'}}{(\sqrt{2s})^{n+n'}}
\sqrt{\frac{1}{(2s-n-q)!(2s-n'-q)!n!n'!}}.
\end{equation}

At this stage, we have the necessary ingredients to study the
behavior of the linear entropy $S$, which is rich. This is because
$S$ is actually depending on three parameters: the truncation $s$ of
the Weyl–Heisenberg algebra, the levels of excitation $|z|$ of the
input coherent states and the reflection coefficient $r$ of the beam
splitter device. We first plot the entanglement against  $r^2$ for
different finite dimensional Hilbert (the dimension is $2s+1$) for
coherent states having an excitation level $z = 5$. The results are
summarized in Figure 1:

\begin{center}
\includegraphics[width=4in]{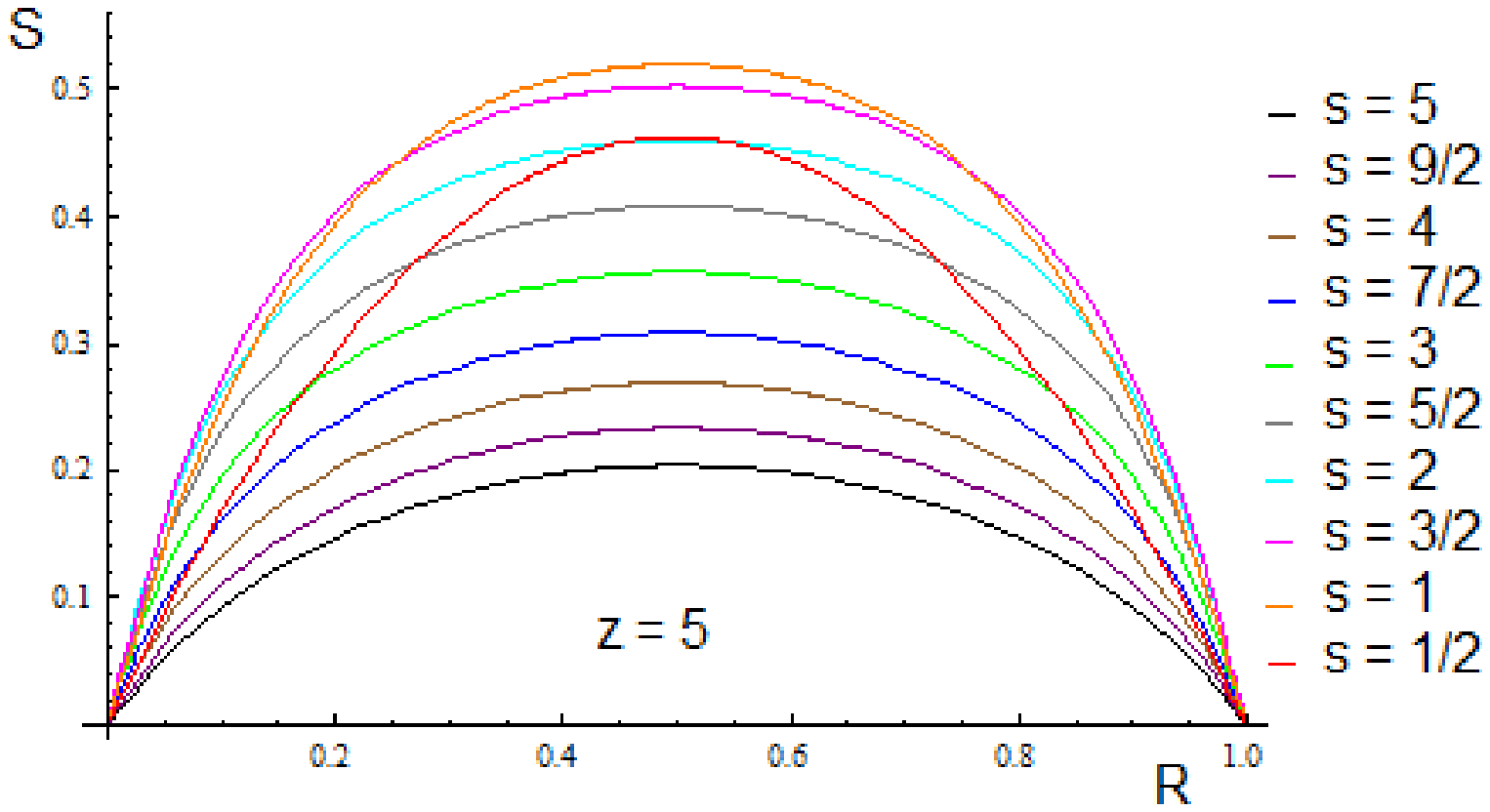}\\
{\sf Figure 1:  {Linear entropy of coherent states as a function of
the square of reflection coefficient $R = r^2$ for the level of
excitation $z =5$.}}
\end{center}
It is clear that the maximum degree of entanglement is reach when
one uses a 50:50 beam splitter (i.e. $ r = 1/\sqrt 2$) as expected.
This is independent of the dimension of the Fock space. It is
remarkable that for $ r = 1/\sqrt 2 $, the linear entropy is maximal
for $s = 1$, which corresponds to qutrits system. The behavior of
the entropy as function of the truncation level of the
Weyl--Heisenberg algebra is represented in Figures 2. We consider
separately coherent states with weak and strong excitation level
passing throughout a 50:50 beam splitter.
\begin{center}
\includegraphics[width=4in]{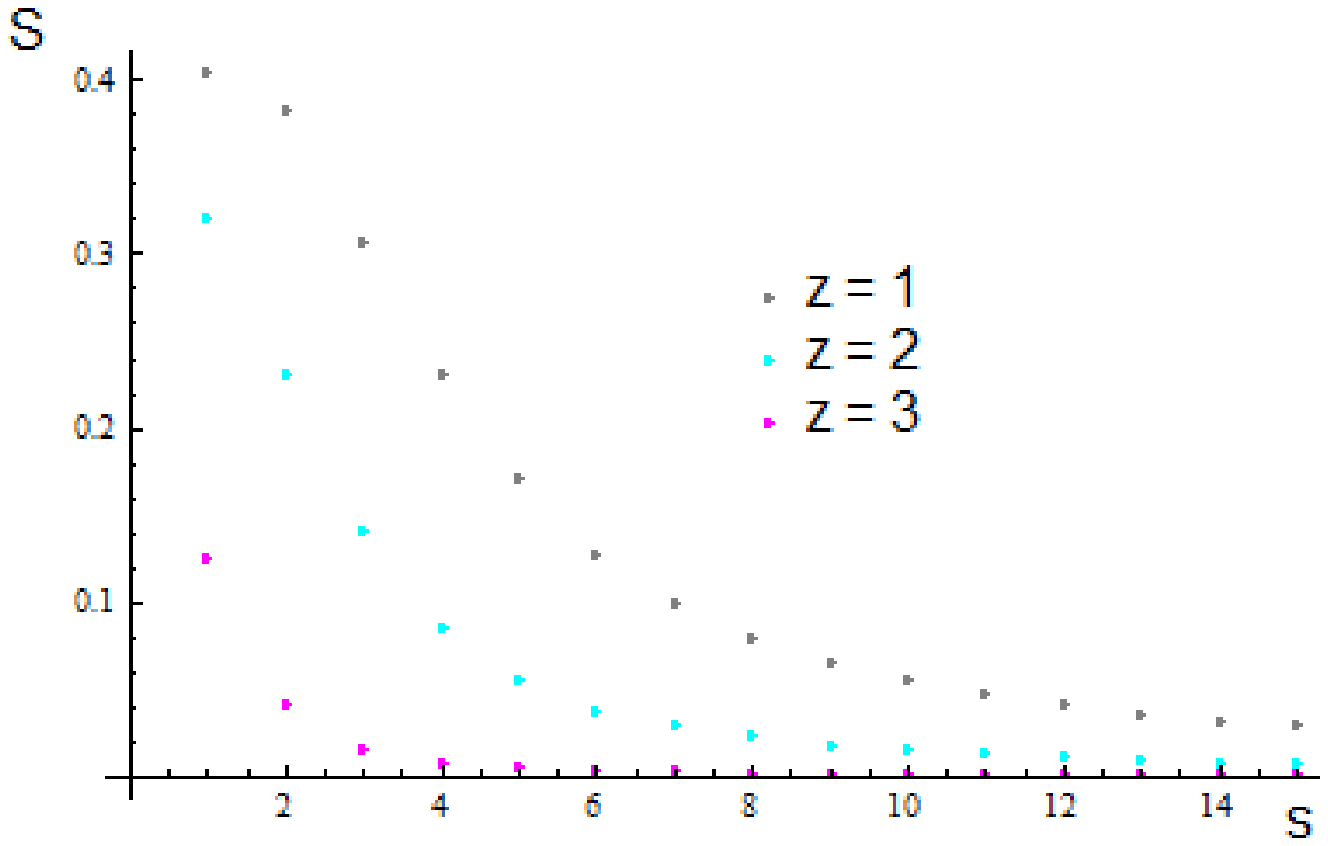}\\
{\sf Figure 2.a:  {Linear entropy  as a function of  $s$ for weak
coherent state excitation.}}
\end{center}
In Figure 2.a, it is clearly shown that for small values of the
variable $z$, the linear entropy decreases quickly and goes to zero
with increasing $s$. This agrees with the fact that for $s$ large,
the linear entropy must vanishes as the coherent states (\ref{cs})
go to Glauber ones for ordinary harmonic oscillator, which does not
exhibit entanglement after passing throughout a 50:50 beam splitter.
In the case of stronger excitation, the linear entropy behaves
differently as it can be seen from Figure 2.b.

\begin{center}
\includegraphics[width=4in]{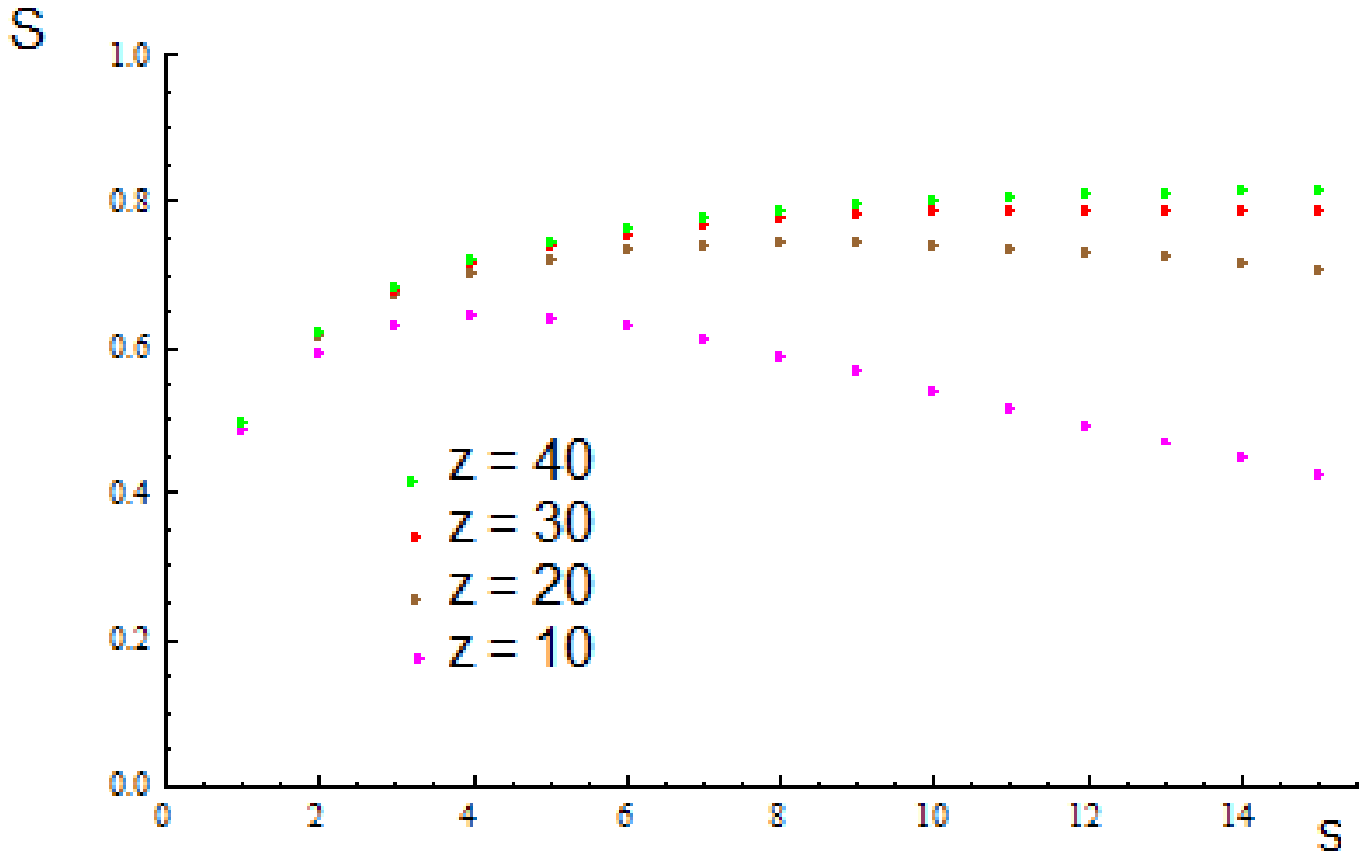}\\
{\sf Figure 2.b:  {Linear entropy  as a function of  $s$ for strong
coherent state excitation.}}
\end{center}
For instance, for $z = 10$, the entropy undergoes an initial
increase followed by a slower decrease. It follows that for strong
excitation levels of the coherent states, the linear entropy will
reach zero for very large value of $s$. To understand the behavior
of the linear entropy as function of the excitation level of the
coherent state $z$, we give the Figure 3:

\begin{center}
\includegraphics[width=4in]{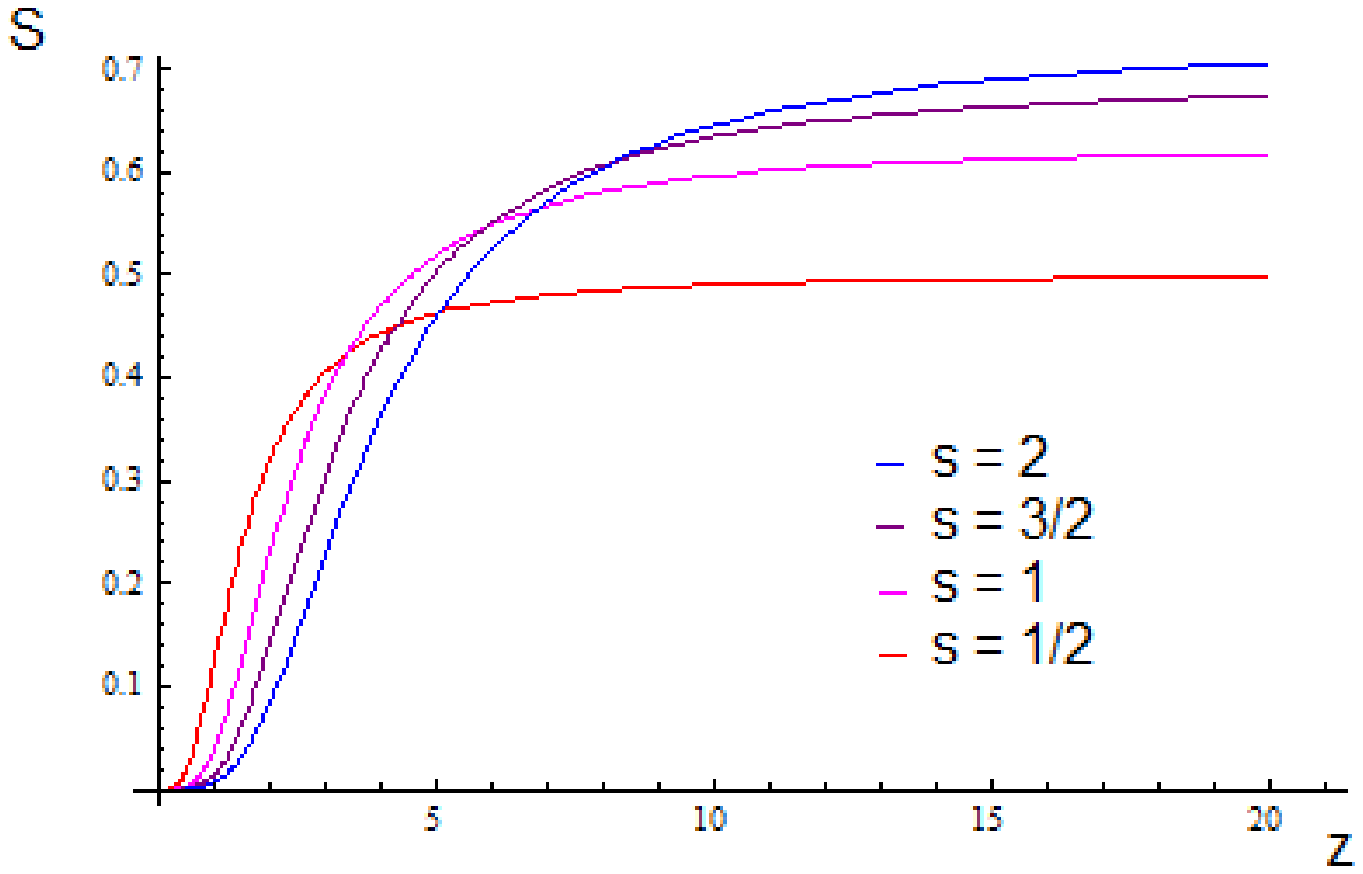}\\
{\sf Figure 3:  {Linear entropy  as a function of  coherent state
excitation.}}
\end{center}
In this Figure, we plot $S$ as function of $z$ for different values
of $s$. It reflects that for weak excitation levels $ z \leq 5$, the
linear entropy increases rapidly and for higher level excitations it
increases slowly. This explains why, for strong level excitation,
 the linear entropy does not go  to zero for higher $s$ so quickly as in the case of systems with small
number of states. The separability of highly excited coherent
states, passing throughout a 50:50 beam splitter, occurs for quantum
systems with sufficiently large number of states (the dimension of
the corresponding Hilbert space is very large).

Two main features were observed through the analysis of this
subsection. The first is that the entanglement of the output states
depends heavily on $|z|$, more than $s$. For weak excitation levels
$z$ of the coherent states, the linear entropy decreases quickly
(Figure 2.a) as $s$ increases. This changes drastically for strong
excitation levels and the entanglement initially increases with $s$
(Figure 2.b). The second interesting feature is that for large
$|z|$, the entanglement, after an initial increase, decreases slowly
as we vary $s$. This decline is very dependent on $|z|$ and is lower
for larger $|z|$. It follows that to see coherent states (with s
large) near zero entanglement, one would need a system of higher
excitation levels. In our numerics we considered  only $s$ such as $
s \leq 15$, but we may see a significantly low resultant
entanglement for large $s$ but with coherent states of extremely
high excitation levels. This reflects the resistance of the
bipartite entanglement of coherent states when passed through a beam
splitter.

\subsection{ Tripartite entanglement induced by two beam splitters}

To investigate the tripartite entanglement,  we now  consider the
action of the unitary operator
\begin{equation}
{\cal B}_{2} = {\cal B}_{2,3}(\theta_2) {\cal B}_{1,2}(\theta_1)
\end{equation}
on the coherent state (\ref{cs}). The operator ${\cal B}_{2}$ is
obtained from (\ref{operatorB}) by setting $k=2$. This situation is
interesting because it allows us to study bipartite as well three
particle entanglement. The input states are of the form $\vert z; 0;
0 \rangle$ where the first mode is prepared in a coherent state
$\vert z \rangle$ and the second and third modes are in their vacuum
state. The action of  ${\cal B}_{2}$ on the state $\vert z , 0 ,
0\rangle$ can be evaluated as follows. Indedd, using (\ref{suncs}),
one verifies
\begin{equation}
{\cal B}_{2} \vert n , 0 , 0\rangle =  \sum_{p=0}^{n} {\sqrt
\frac{n!}{p!(n-p)!}} t_1^p (ir_1)^{n-p} \sum_{p'=0}^{n-p} {\sqrt
\frac{(n-p)!}{p'!(n-p-p')!}}
 t_2^{n-p} (ir_2)^{n-p-p'} \vert p , n-p , n-p-p'\rangle
\end{equation}
from which one obtains
\begin{equation}
{\cal B}_{2} \vert z , 0 , 0\rangle \equiv \vert {\rm output}
\rangle =  \sum_{n=0}^{2s} \sum_{p=0}^{n} \sum_{p'=0}^{n-p}
a(n,p,p')  \vert p , p' , n-p-p'\rangle \label{output}
\end{equation}
where
\begin{equation}
a (n,p,p') = {\cal N} C_{n} z^n{\sqrt \frac{n!}{p!(n-p)!}} t_1^p
(ir_1)^{n-p}  {\sqrt \frac{(n-p)!}{p'!(n-p-p')!}} t_2^{p'}
(ir_2)^{n-p-p'}. \label{coefa}
\end{equation}

The output state is a pure state of three qudits (1, 2 and 3) system
in the basis $\{ \vert n_1 , n_2 , n_3 \rangle \}$. Then, according
to \cite{Ma}, for the pure state $\vert {\rm output} \rangle$  given
by (\ref{output}) defined on ${\cal F}\otimes{\cal F}\otimes{\cal
F}$ (dim ${\cal F} = 2s+1$),
 the concurrence has the form
\begin{equation}
C (\vert {\rm output} \rangle) = \sqrt{\frac{2s+1}{12s}\left[ 3 -
\left({\rm Tr}\rho_1^2+{\rm Tr}\rho_2^2+{\rm
Tr}\rho_3^2\right)\right]}. \label{concurence}
\end{equation}
where
\begin{equation}
\rho_1 =  {\rm Tr}_{2,3} \rho,  \qquad \rho_2 =  {\rm Tr}_{1,3}
\rho, \qquad \rho_3 =  {\rm Tr}_{1,2} \rho \label{rho1rho2rho3}
\end{equation}
and $\rho = \vert {\rm output} \rangle \langle {\rm output} \vert$.
This formula provides us with a measure of the tripartite
entanglement of the coherent state (\ref{cs}) after passing through
two beam splitters. We can interpret the right hand of the equation
(\ref{concurence}) as follows. If we regard the pair of oscillators
(23) as a single subsystem, it makes sense to talk about the
bipartite entanglement between the first oscillator system 1 and the
pair (23). In this respect, one can treat 1 and (23) as two systems
and
use the linear entropy 
as measure of the degree of entanglement between them. In the same
manner, one can define the bipartite entanglement between the
subsystems 2 and (13) and the pair 3 and (12). The linear entropies
in each case are given by
\begin{equation}
S_1 = 1 - {\rm Tr} \rho_1^2, \qquad  S_2 = 1 - {\rm Tr} \rho_2^2,
\qquad  S_3 = 1 - {\rm Tr} \rho_3^2. \label{S1S2S3}
\end{equation}
It follows that the concurrence (\ref{concurence}) can be expressed
in terms of the sum of $S = S_1 + S_2 + S_3$. This reflects that the
tripartite entanglement is related to the degree of bipartite
entanglement between the different subcomponents of the system under
consideration.

Figures 4, 5 and 6 present the behavior of the entropies $S_1$,
$S_2$ and $S_3$ (\ref{S1S2S3}) in terms of the square of the
reflection coefficients $r_1$ and $r_2$ of the beam splitters acting
in the coherent state (\ref{cs}). We focus on qutrits system, i.e.
$s=1$ and  take the level of excitation of the input coherent state
as $ z= 5$.

\begin{center}
  \includegraphics[width=4in]{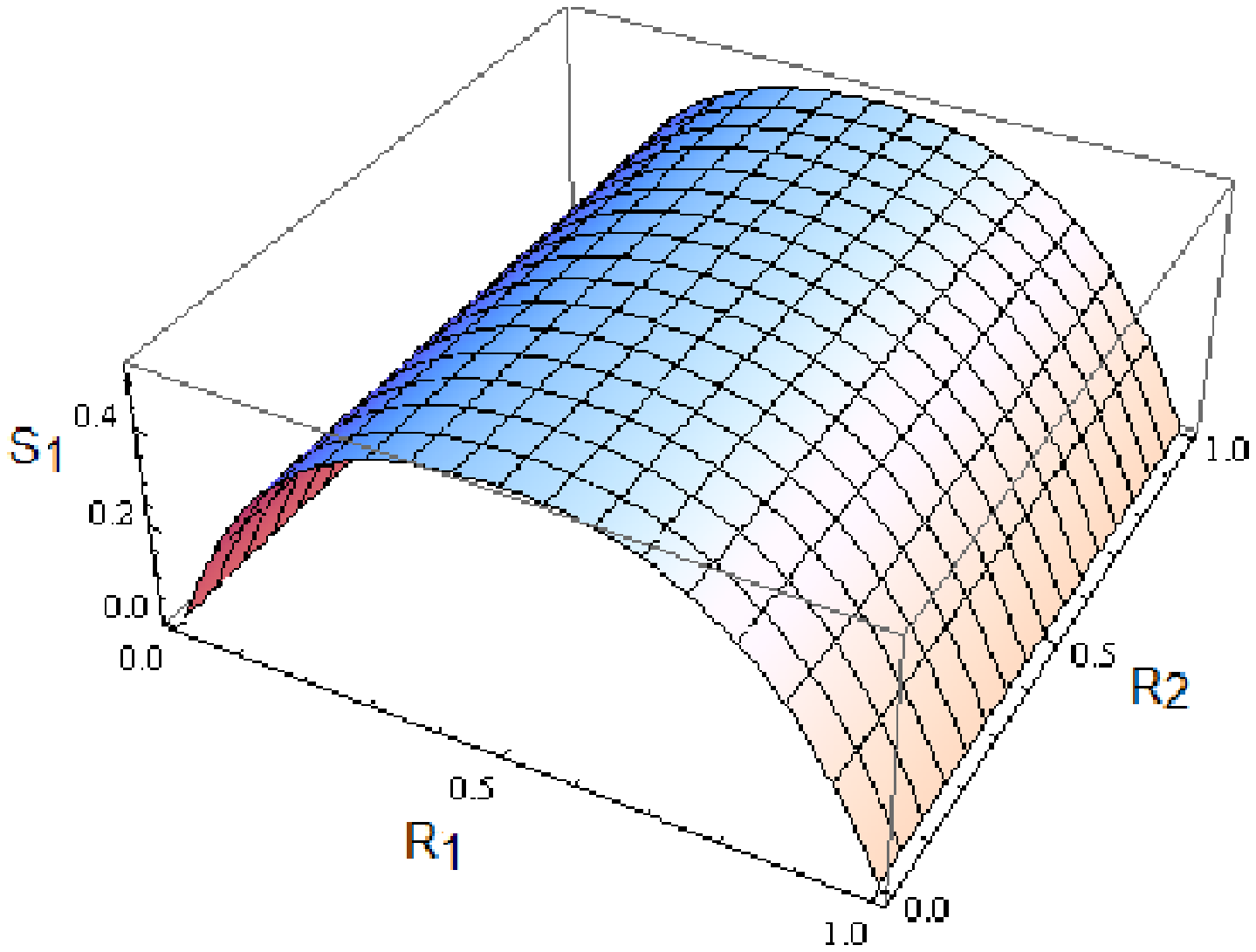}\\
{\sf Figure 4:  {Linear entropy $S_1$ versus $R_1 = r^2_1$ and $R_2
= r^2_2$ for $(s=1, z=5)$.}}
\end{center}
The linear entropy $S_1$  measuring the degree of bipartite
entanglement between the parts 1 and (23) viewed as a single
subsystem is plotted in Figure 4. It is important to note that $S_1$
is maximal for $r_1 = 1/\sqrt 2$ and $r_2$-independent. This result
can be easily shown analytically using the equations (\ref{output}),
(\ref{rho1rho2rho3}) and (\ref{S1S2S3}).  This is not the case for
the entropy $S_2$ as it is shown by Figure 5:

\begin{center}
  \includegraphics[width=4in]{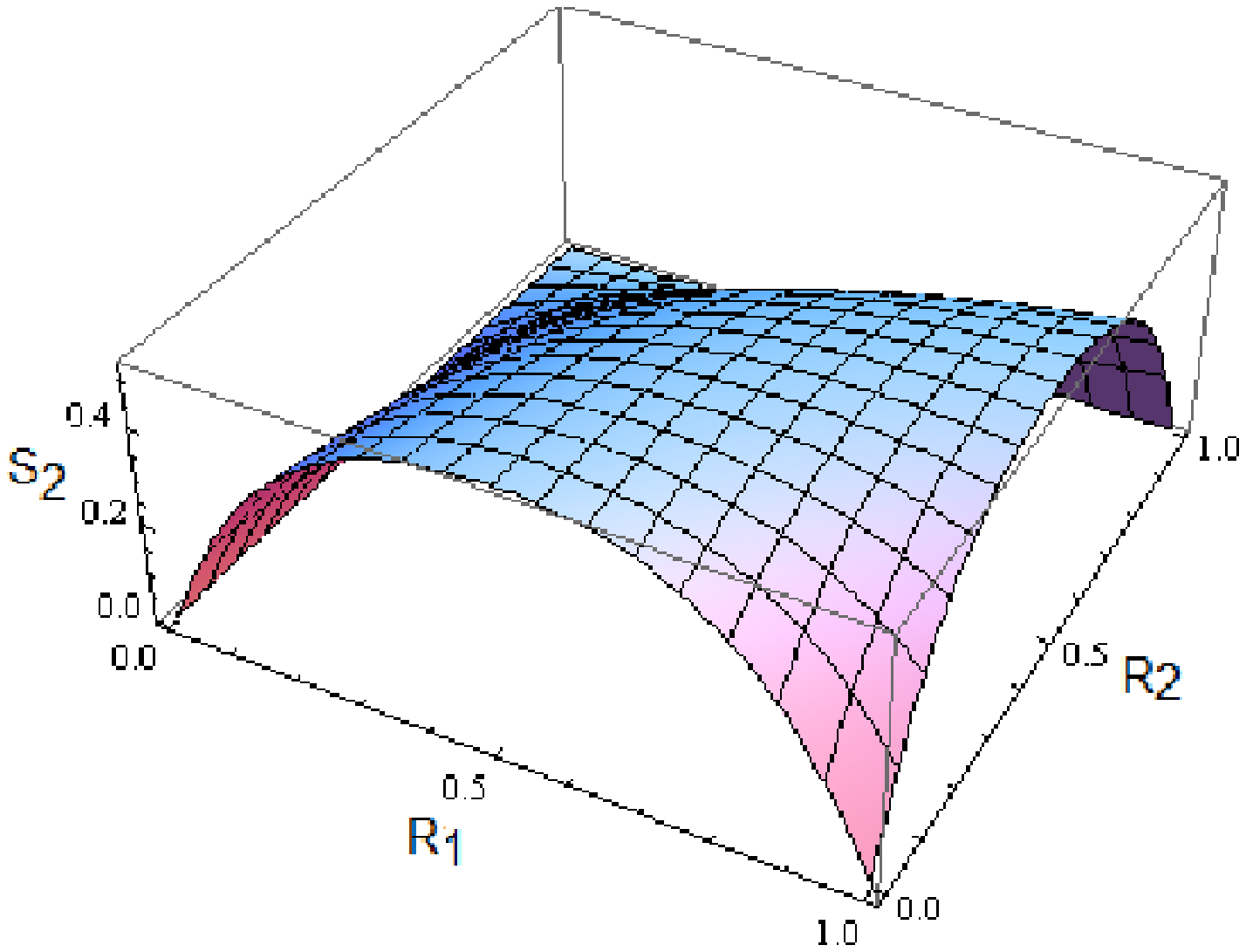}\\
{\sf Figure 5:  {Linear entropy $S_2$ versus $R_1 = r_1^2$ and $R_2
= r_2^2$ for $(s=1, z=5)$.}}
\end{center}
The quantity $S_2$ is associated  with the linear entropy measuring
the degree of bipartite entanglement between the subsystems $2$ and
$(13)$. The entropy $S_2$ is maximal for $(r_1 = 1/\sqrt 2 , r_2 =
0)$ or $(r_1 = 1, r_2 =  1/\sqrt 2)$ and vanishes for $r_1 = 0$ or
$r_2 = 1$. Indeed, using the equation (\ref{output}), one can simply
check that the output state is a tensorial product of a wave
function corresponding to the subsystem $(13)$ and the vacuum $\vert
0 \rangle$ of the bosonic mode $2$. Consequently, the linear
entropy, associated with entanglement of the subsystems $2$ and
$(13)$, vanishes in agreement with the results of Figure 5.
Similarly, we plot in Figure 6 the linear entropy $S_3$ which gives
the amount of entanglement between the subsystems $(12)$ and $3$.

\begin{center}
  \includegraphics[width=4in]{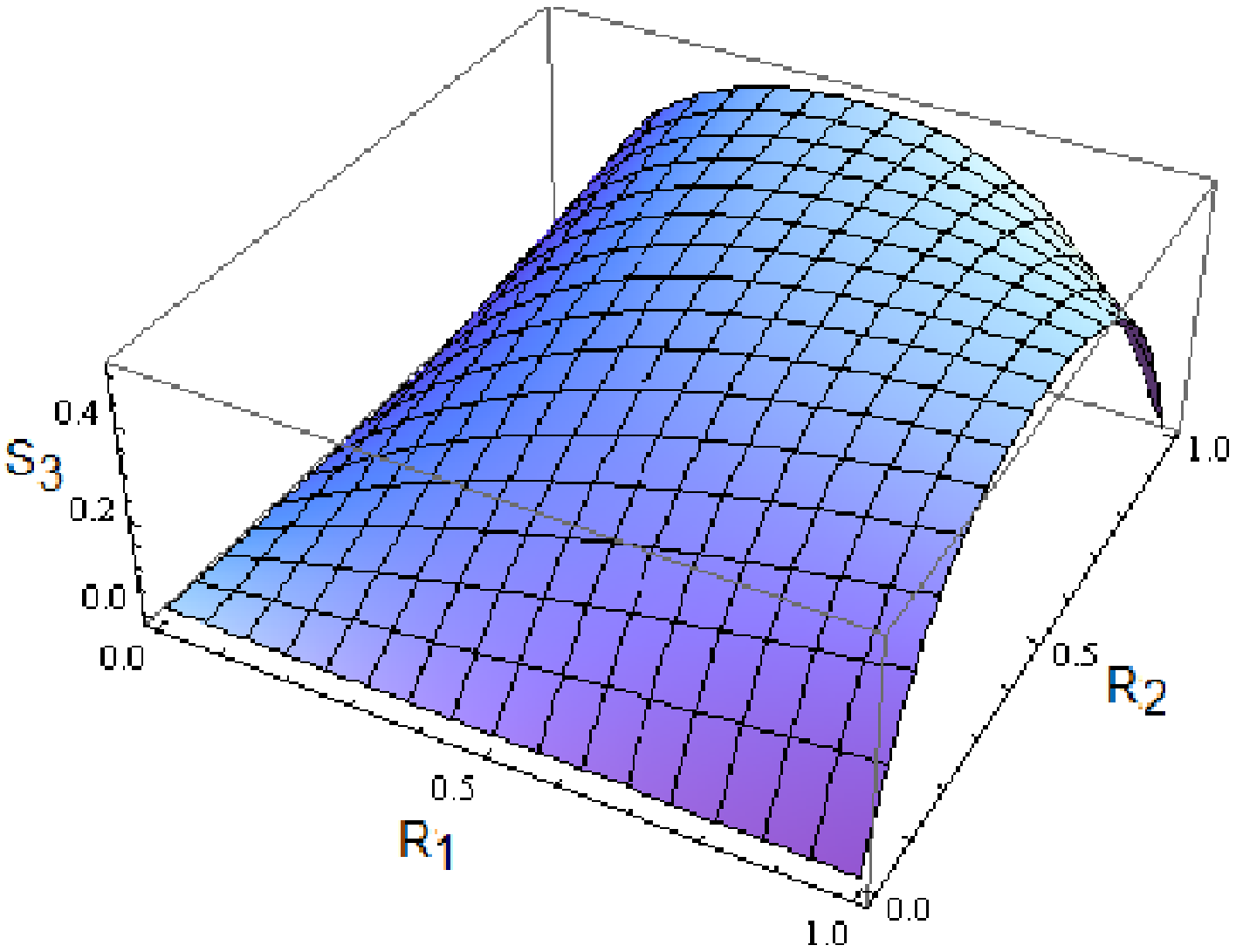}\\
{\sf Figure 6:  {Linear entropy $S_3$ versus $R_1 = r_1^2$ and $R_2
= r_2^2$ for $(s=1, z=5)$.}}
\end{center}
The maximal value of the linear entropy $S_3$ is reached  for $(r_1
= 1/\sqrt 2 , r_2 = 1)$ or  $(r_1 = 1 , r_2 = 1/\sqrt 2)$ and
 vanishes for $r_1 = 0$ or $r_2 = 0$. This can be
easily verified using the equation (\ref{output}). Indeed, for $r_1
= 0$, the output state is separable because the action of the beam
splitters leaves invariant the state $\vert z, 0, 0 \rangle$.
Similarly, for $r_2 = 0$, the third part of the system, initially in
the vacuum state, remains unchanged when passed throughout the beam
splitters device and the output state is
completely disentangled.\\

At this stage, one can use the linear entropies $S_1$, $S_2$ and
$S_3$  to study the tripartite entanglement of the output state
under consideration. In fact, from (\ref{concurence}), it is easily
seen that the square of the tripartite concurrence is proportional
to the sum of the entropies $S_1$, $S_2$ and $S_3$. This quantity is
plotted in Figure 7:
\begin{center}
  \includegraphics[width=4in]{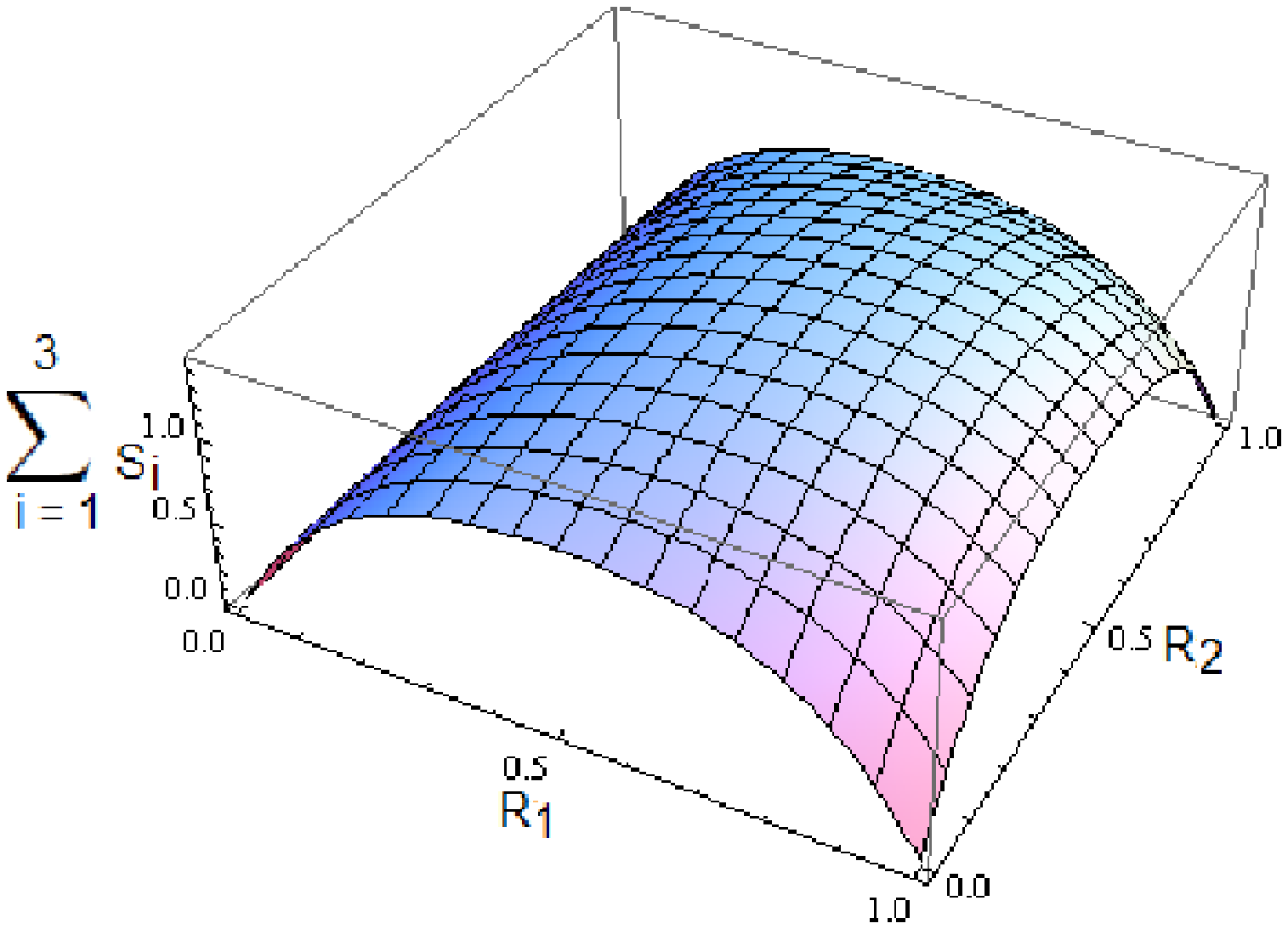}\\
{\sf Figure 7:  {The total Linear entropy $S = S_1+S_2+S_3$ as
functions of $R_1 = r^2_1$ and $R_1 = r^2_2$ for $(s=1,z=5)$.}}
\end{center}
We notice that for coherent states of a qutrit system ($s =1$) with
$z=5$, the maximal tripartite entanglement is obtained for $r_1 =
r_2 = 1/\sqrt 2$. It is interesting to mention that in the
particular cases where $(r_1 = 0, r_2 = 0)$, $(r_1 = 1, r_2 = 0)$,
$(r_1 = 0 , r_2 = 1)$ and $(r_1 = 1, r_2 = 1)$, the output states
are given by $\vert z, 0, 0 \rangle$, $\vert 0, iz, 0 \rangle$,
$\vert z, 0, 0 \rangle$ and $\vert 0, 0, -z \rangle$, respectively.
These states, derived from the equation (\ref{output}), are clearly
separable. This agrees with  Figure 7 from which one can see that
the entropy $S$ vanishes only in these particular four situations.
In this subsection, we only considered qutrits. This study can be
extended to any  qudit system ($s$ arbitrary). It should be noticed
that for $s$ large and high excitation levels $z$, no entanglement
can be generated at the output. This is essentially due, as we
discussed in the previous subsection, to the fact that in this limit
all the bipartite entanglement in the system vanishes and hence no
tripartite entanglement can be generated. Indeed, for $s$ large the
coherent states (\ref{cs}) reduces to Glauber ones. In this case,
the action of two 50:50 beam splitters (i.e. $ir_1 = ir_2 =
\frac{1}{\sqrt 2}$), with a Glauber coherent states $\vert z )$
incident in one port and a two particle vacuum $\vert 0 , 0 \rangle
$ on the other ports, reduces to
\begin{equation}
\vert z )~ \vert 0 \rangle ~\vert 0 \rangle \longrightarrow \vert
z/{\sqrt 2} )~ \vert i z/2 ) ~\vert - z/2 )
\end{equation}
where the notation $\vert \alpha )$ stands for  the Glauber coherent
states, such as
$$ \vert \alpha ) = e^{-|\alpha|^2/2} \sum_{n=0}^{\infty} \frac{\alpha^n}{n!} \vert n \rangle. $$
This shows that in the limiting case $s \to \infty$, the resulting
output state is disentangled. The results presented in this section
can be extended in different directions to investigate many features
of entangled coherent states passing throughout a quantum network of
beam splitters.

\section{\bf Conclusion}


Using the linear entropy as measure of the bipartite entanglement,
we have investigated the bipartite entanglement of the coherent
states of a truncated Weyl--Heisenberg algebra. This algebra
constitutes an alternative way to describe quantized electromagnetic
field with finite dimensional Hilbert spaces instead of the usual
truncated harmonic oscillator introduced by Pegg and Barnett
\cite{Pegg}. Subsequently, We have discussed the use of a quantum
network of $k$ beam splitters as experimental device  to generate
easily the coherent states associated with $SU(k+1)$ algebra as well
as the entangled photonic states.

We have focused on network involving one or two beam splitters. In
the first case, this device is used to investigate the degree of
entanglement of coherent state associated with the truncated
harmonic oscillator. We have shown that, for weak level of
excitation, the linear entropy goes faster to zero for $s$
increasing. For stronger excitation levels, the entropy decreases
slowly as the Fock space dimension increases. The second case
involves two beams splitters. In this case taking advantage from the
relation of the tripartite concurrence  and the partial linear
entropies (see equation (\ref{concurence})), we have discussed the
bipartite as well tripartite entanglement of a coherent states
passing throughout a network of two beam splitters.  Moreover, we
have shown that the maximal tripartite entanglement is reached when
the beam splitters are 50:50.


\section{Acknowledgments}

MD would like to express his thanks to Max Planck Institute for
Physics of Complex Systems (Dresden-Germany) where this work was
done. AJ $\&$ EBC thank the partial support of the Saudi Center for
Theoretical Physics (SCTP) of Dhahran $\&$ King Faisal University
(KFU).


\end{document}